\documentclass[11pt]{article}
\usepackage[utf8]{inputenc}
\usepackage{longtable}
\usepackage[singlelinecheck=off]{caption}
\usepackage{subcaption}
\usepackage{graphicx}
\usepackage{placeins}
\usepackage{booktabs}
\usepackage{float}
\usepackage{comment}
\usepackage[T2A]{fontenc}
\PassOptionsToPackage{hyphens}{url}\usepackage{hyperref}

\usepackage{amsmath, bbm, amssymb}
\usepackage[dvipsnames,table,xcdraw,svgnames]{xcolor}
\usepackage[hmargin=0.8in,vmargin={1.1in,1.1in}]{geometry}
\usepackage{hyperref}

\usepackage{multicol}
\hypersetup{
   colorlinks=true,
    linkcolor=blue,
   citecolor=blue,      
   urlcolor=blue,
}
\usepackage{natbib}
\setcitestyle{citesep={;}}
\usepackage[font={footnotesize}]{caption}
\usepackage{url} 

\usepackage[onehalfspacing]{setspace}

\usepackage{parskip}
\usepackage{array}
\newcolumntype{P}[1]{>{\centering\arraybackslash}p{#1}}

\title{
Are Large Language Models a Threat to Digital Public Goods? Evidence from Activity on Stack Overflow}

\author{Maria del Rio-Chanona $^{1,2}$, Nadzeya Laurentsyeva $^{3}$, and Johannes Wachs $^{4,5,1,*}$ \\
\footnotesize{$^{1}$ Complexity Science Hub, Vienna. }\\
\footnotesize{$^{2}$ Harvard Kennedy School}\\
\footnotesize{$^{3}$ Faculty of Economics, LMU Munich}\\
\footnotesize{$^{4}$ Corvinus University of Budapest}\\
\footnotesize{$^{5}$ Centre for Economic and Regional Studies, Hungary}
\thanks{Direct correspondence to johannes.wachs@uni-corvinus.hu}
}

\date{2023-07-14}

\begin{document}
\maketitle

\begin{abstract}

Large language models like ChatGPT efficiently provide users with information about various topics, presenting a potential substitute for searching the web and asking people for help online. But since users interact privately with the model, these models may drastically reduce the amount of publicly available human-generated data and knowledge resources. This substitution can present a significant problem in securing training data for future models. In this work, we investigate how the release of ChatGPT changed human-generated open data on the web by analyzing the activity on Stack Overflow, the leading online Q\&A platform for computer programming.
We find that relative to its Russian and Chinese counterparts, where access to ChatGPT is limited, and to similar forums for mathematics, where ChatGPT is less capable, activity on Stack Overflow significantly decreased. A difference-in-differences model estimates a 16\% decrease in weekly posts on Stack Overflow. This effect increases in magnitude over time, and is larger for posts related to the most widely used programming languages. Posts made after ChatGPT get similar voting scores than before, suggesting that ChatGPT is not merely displacing duplicate or low-quality content. These results suggest that more users are adopting large language models to answer questions and they are better substitutes for Stack Overflow for languages for which they have more training data. Using models like ChatGPT may be more efficient for solving certain programming problems, but its widespread adoption and the resulting shift away from public exchange on the web will limit the open data people and models can learn from in the future.

\end{abstract}

\clearpage

\section{Introduction}

Over the last thirty years, humans have constructed a vast library of information on the web. Using powerful search engines anyone with an internet connection can access valuable information from online knowledge repositories like Wikipedia, Stack Overflow, and Reddit. New content and discussions posted online are quickly integrated into this ever-growing ecosystem, becoming digital public goods used by people all around the world to learn new technologies and solve their problems \citep{hess2003ideas,henzinger2004extracting,lemmerich2019world,piccardi2021value}. 

More recently, these public goods have been used to train artificial intelligence (AI) systems, in particular, large language models (LLMs) \citep{vaswani2017attention}. For example, the LLM ChatGPT \citep{openai2023gpt4} answers user questions by summarizing the information contained in these repositories. The remarkable effectiveness of ChatGPT is reflected in its quick adoption \citep{teubner2023welcome} and application across diverse fields including auditing \citep{gu2023artificial}, astronomy \citep{smith2023astronomia}, medicine \citep{kanjee2023accuracy}, and chemistry \citep{guo2023indeed}. Randomized control trials show that using LLMs significantly boosts productivity in computer programming, professional writing, and customer support tasks \citep{peng2023impact,noy2023experimental,brynjolfsson2023generative}. Indeed, the widely reported successes of LLMs like ChatGPT suggest that we will observe a significant change in how people search for, create and share information online.

Ironically, if LLMs like ChatGPT present substitute traditional ways of searching and interrogating the web, then they will displace the very human behavior that generated their original training data. User interactions with ChatGPT are the exclusive property of OpenAI, its creator. Only OpenAI will be able to learn from the information contained in these interactions. As people begin to use LLMs instead of online knowledge repositories to find information, contributions to these repositories will likely decrease, diminishing the quantity and quality of these digital public goods. While such a shift would have significant social and economic implications, we have little evidence on whether people are actually substituting their consumption and creation of digital public goods with ChatGPT. 

The aim of this paper is to evaluate the impact of LLMs on the generation of open data on question-and-answer (Q\&A) platforms. Since LLMs perform relatively well on software programming tasks~\citep{peng2023impact}, we study Stack Overflow, the largest online Q\&A platform for software development and programming. We present three results. First, we examine whether the release of ChatGPT has decreased the volume of posts, i.e. questions and answers, posted on the platform. We measure the overall effect of ChatGPT's release on Stack Overflow activity using a difference-in-differences model. We compare the weekly posting activity on Stack Overflow against that of four comparable Q\&A platforms. These counterfactual platforms are less likely to be affected by ChatGPT either because their users are less able to access ChatGPT or because ChatGPT performs poorly in questions discussed on those platforms. We find that posting activity on Stack Overflow decreased by about $16\%$ following the release of ChatGPT, increasing over time to around $25\%$ within six months.

Second, we investigate whether ChatGPT is simply displacing simpler or lower quality posts on Stack Overflow. To do so, we use data on up- and downvotes, simple forms of social feedback provided by other users to rate posts. We observe no change in the votes posts receive on Stack Overflow since the release of ChatGPT. This finding suggests that ChatGPT is displacing a wide variety of Stack Overflow posts, including high-quality content. 

Third, we study the heterogeneity of the impact of ChatGPT across different programming languages discussed on Stack Overflow. We test for these heterogeneities using an event study design. We observe that posting activity in some languages like Python and Javascript has decreased significantly more than the global site average. Using data on programming language popularity on GitHub, we find that the most widely used languages tend to have larger relative declines in posting activity.

Our analysis points to several significant implications for the sustainability of the current AI ecosystem. The first is that the decreased production of open data will limit the training of future models \citep{villalobos2022will}. LLM-generated content itself is an ineffective substitute for training data generated by humans for the purpose of training new models \citep{gudibande2023false,shumailov2023recursion, alemohammad2023selfconsuming}. One analogy is that training an LLM on LLM-generated content is like making a photocopy of a photocopy, providing successively less satisfying results \citep{chiang2023chatgpt}. And while human feedback to LLMs may facilitate continued learning, such feedback remains private information. This suggests a second issue: ChatGPT's initial advantage can compound if it effectively learns from its interactions with users while simultaneously crowding out the generation of new open data \citep{arthur1989competing}. More broadly, a shift from open data to a more closed web will likely have significant second-order impacts on the digital economy and how we access and share information.

The rest of the paper is organized as follows. We introduce our empirical set-up, including the data and models used in our analysis, in Section~\ref{sec:data_methods}. Section~\ref{sec:results} presents our results. In Section~\ref{sec:discussion}, we discuss their implications. We argue that our findings of a significant decline in activity on Stack Overflow following the release of ChatGPT have important implications for the training of future language models, competition in the artificial intelligence sector, the provision of digital public goods, and how humans seek and share information. 

\section{Data and Methods}\label{sec:data_methods}
\subsection{Stack Exchange and Segmentfault data}\label{sec:data_main_text}
To understand the effect ChatGPT can have on digital public goods, we compare the change in Stack Overflow's activity with the activity on a set of similar platforms. These platforms are similar to Stack Overflow in that they are technical Q\&A platforms, but are less prone to substitution by ChatGPT given their focus or target group. Specifically, we focus on the Stack Exchange platforms Mathematics and Math Overflow and on the Russian-language version of Stack Overflow. We also examine a Chinese-language Q\&A platform on computer programming called Segmentfault.

Mathematics and Math Overflow focus on university- and research-level mathematics questions respectively. We consider these sites to be less susceptible to replacement by ChatGPT given that, during our study's period of observation, the free-tier version of ChatGPT performed poorly (0-20th percentile) on advanced high-school mathematics exams \citep{openai2023gpt4}, and was therefore unlikely to serve as a suitable alternative to these platforms.

The Russian Stack Overflow and the Chinese Segmentfault have the same scope as Stack Overflow, but target users located in Russia and China, respectively. We consider these platforms to be less affected by ChatGPT given that ChatGPT is officially unavailable in the Russian Federation, Belarus, Russian-occupied Ukrainian territory, and the People's Republic of China. Although people in these places can and do access ChatGPT via VPNs \citep{kreitmeir2023unintended}, such barriers still represent a hurdle to widespread fast adoption.

We extract all posts (questions or answers) on Stack Overflow, Mathematics, Math Overflow, and Russian Stack Overflow from their launch to early June 2023 using \url{https://archive.org/details/stackexchange}. We scraped the data from Segmentfault directly. Our dataset comprises 58 million posts on Stack Overflow, over 900 thousand posts for the Russian-language version of Stack Overflow, 3.5 million posts on Mathematics Stack Exchange, 300 thousand posts for Math Overflow, and about 300 thousand for Segmentfault. We focus our analysis on data from January 2019 to June 2023, noting that our findings are robust to alternative time windows.

For each post, our dataset includes the number of votes (up -- positive feedback, or down -- negative feedback) the post received, the author (user), and whether the post is a question or an answer. Furthermore, each post can have up to 5 tags -- predefined labels that summarize the content of the post, for instance, an associated programming language. For more details on the data used, we refer the reader to section~\ref{sec:methods}. From this point forward, we will refer to Mathematics, Math Overflow, Russian Stack Overflow, and Segmentfault, along with their corresponding posts, as the counterfactual platforms and posts.

\subsection{Models}\label{sec:model_main}

\paragraph{Difference-in-differences}
We estimate the effect of ChatGPT for posting activity on Stack Overflow using a difference-in-differences method with four counterfactual platforms. We aggregate posting data at platform- and week-level and fit a regression model using ordinary least squares (OLS):
\begin{equation}
IHS(Posts_{p,t}) = \alpha_{p} +\lambda_{t} + \beta \times Treated_{p,t} + 
\sum_{p\in P} \theta_{p}t 
+ \epsilon_{p,t} 
\end{equation}

\noindent where $Posts_{p,t}$ is the number of posts on platform $p$ in a week $t$, which we transform using the inverse-hyperbolic sine function (IHS)
\citep{burbidge1988alternative}.\footnote{We prefer this transformation because then the coefficient of interest can be roughly interpreted as a percent change in posting activity. The IHS behaves similarly to a natural log transformation for positive values but remains defined for zeroes. Our estimates are qualitatively similar to using log transformation, standardization or raw data.} 
$\alpha_{p}$ are platform fixed effects, $\lambda_{t}$ are time (week) fixed effects, $\theta_{p}$ are platform-specific linear time trends, and $\epsilon_{p,t}$ is the error term. 

The coefficient of interest is $\beta$, which captures the estimated effect of ChatGPT on posting activity on Stack Overflow relative to the less affected platforms: $Treated$ equals one for weeks after the release of ChatGPT (starting November 27, 2022) when the platform $p$ is Stack Overflow and zero otherwise. We report robust standard errors clustered at the monthly level. 

To check the dynamics of the effect and to examine pretrends, we employ a similar specification but instead of $\beta \times Treated_{p,t}$, we use $\sum_t \beta_t \times I(week = t) \times I(platform = StackOverlow)$. We standardize the effects to 0 in the week before the public release of ChatGPT by dropping the indicator for that week from the regression. Separate coefficients for 25 weeks \textit{following} the release of ChatGPT show how the effects of ChatGPT realized over time. Separate coefficients for the first 100 weeks \textit{before} the release allow us to verify that posts on Stack Overflow had evolved similarly to the activity on counterfactual platforms prior to the introduction of ChatGPT. 

The advantage of the difference-in-differences method compared to a simple event study with Stack Overflow data only is that we estimate ChatGPT effects net of possible weekly shocks that are common across the technical Q\&A platforms. For the interpretation of the coefficient, we note that we estimate \textit{relative} change in posting activity on Stack Overflow compared to activity on other platforms before vs. after the release of ChatGPT.  

\paragraph{Event Study}

When analyzing the effect of ChatGPT on activity across programming languages, we can no longer compare data from Stack Overflow with the counterfactual platforms. This is because the tags annotating posts are different between Stack Exchange platforms. Therefore, we study ChatGPT's heterogeneous effects using an event-study specification. For each programming language $i$ (identified by a tag), we model the standardized number of posts in a week $t$ on Stack Overflow by fitting a simple linear time trend with seasonal effects:

\begin{equation}
    \overline{Posts}_{i,t}= \beta_{0} +\beta_{1}(t) + 
\beta_{2}(ChatGPT) + \beta_{3}(t \times ChatGPT) + \eta + \epsilon_{i,t}
\end{equation}
where $t$ is the linear time trend and $\eta$ are seasonal (month of year) fixed effects. $ChatGPT$ equals one if the week $t$ is after the release of ChatGPT and zero otherwise. Coefficient $\beta_2$ captures the change in the intercept and coefficient $\beta_3$ reflects the change in the slope of the time trend following the release of ChatGPT. In the tag-level analysis, we standardize the dependent variable in order to be better able to compare effects across programming languages with different numbers of posts.\footnote{We standardize the number of posts within each tag by subtracting the mean and dividing by the standard deviation. Both statistics are calculated before the release of ChatGPT.} We report HAC standard errors.

\section{Results}\label{sec:results}

\begin{figure}[tbh]
    \centering
    \includegraphics[width=\textwidth]{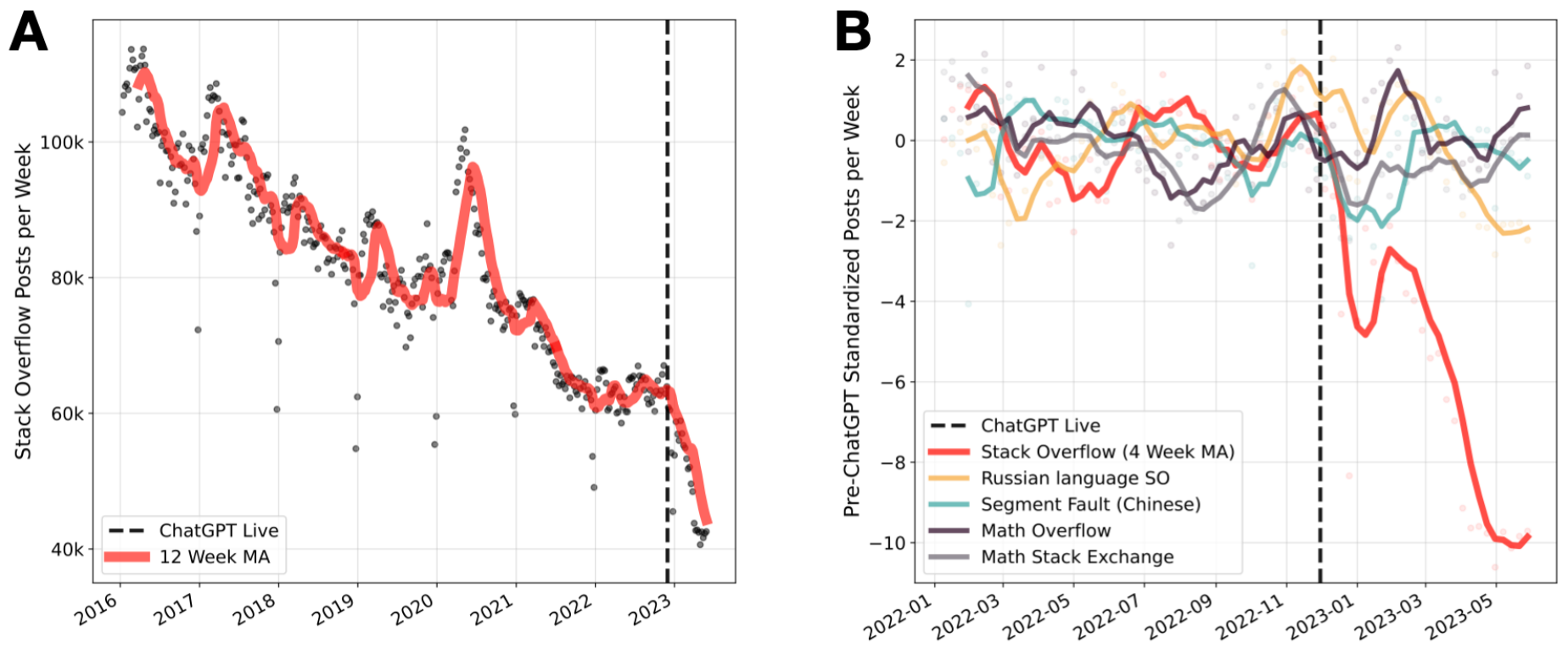}
    \caption{A) Time series of weekly posts to Stack Overflow since early 2016. The number of weekly posts decreases at a rate of about 7,000 posts each year from 2016 to 2022. In the six months after the release of ChatGPT, the weekly posting rate decreases by around 20,000 posts. B) Comparing posts to Stack Overflow, its Russian- and Chinese-language counterparts, and mathematics Q\&A platforms since early 2022. Post counts are standardized by the average and standard deviation of post counts within each platform prior to the release of ChatGPT. Posting activity on Stack Overflow falls significantly more relative to activity on other platforms.}
    \label{fig:combined_figure1}
\end{figure}
\subsection{Decrease in posting activity}

Figure~\ref{fig:combined_figure1}A, shows the evolution of activity on Stack Overflow from January 2016 to June 2023. Up to 2022 there was a gradual decrease in activity from roughly 110,000 to 60,000 posts per week, that is roughly 7,000k posts less per week each year. However, after the release of ChatGPT (November 30th, 2022) posting activity decreased sharply, with the weekly average falling from around 60,000 posts to 40,000 within six months. Compared to the pre-ChatGPT trend, this decrease represents more than five years worth of deceleration in just half a year.

The decrease in activity on Stack Overflow is larger than for similar platforms for which we expect ChatGPT to be a less viable substitute. Figure~\ref{fig:combined_figure1}B shows the standardized posting activity on Stack Overflow, the Russian- and Chinese-language counterparts of Stack Overflow, and two mathematics Q\&A platforms. We standardize posting activity by the average and standard deviation of post counts within each platform prior to the release of ChatGPT.

Figure~\ref{fig:combined_figure1}B shows that Stack Overflow activity deviates markedly from activity on the other platforms after the release of ChatGPT. The plot visualizes the standardized posting activity within each platform since early 2022. Smoothed weekly activity varies between plus and minus two standard deviations for all platforms for most of 2022. Events, such as the Chinese New Year and other holidays and the start of the Russian invasion of Ukraine, are visible. Following the release of ChatGPT, we observe a significant decline in activity on Stack Overflow.

\begin{figure}[H]
    \centering
    \includegraphics[width=\textwidth]{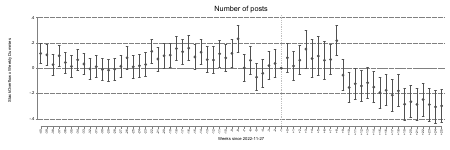}
    \caption{Difference-in-differences analysis for posting activities. The dashed line marks November 30, 2023 the release date of ChatGPT. Eight weeks after its introduction, we observe a steady decline in the activity of Stack Overflow. The plotted coefficients correspond to the interaction between a weekly dummy and posting on Stack Overflow. The coefficients are normalized to that in the week before the release of ChatGPT. The reported confidence intervals are at 95\%. The regression comprises platform fixed effects, week fixed effects, and platform-specific linear time-trends.}
    \label{fig:interaction_plot}
\end{figure}

We report the estimated effect of our difference-in-differences model in Table~\ref{tab:diff_in_diff} and visualize the weekly estimates of the relative change in the Stack Overflow activity in Figure~\ref{fig:interaction_plot}. Table~\ref{tab:diff_in_diff} indicates that ChatGPT decreased posting activity on Stack Overflow by 15.6\% ($1-e^{-0.17}$). These results are robust to changes in the controls and starting point of the data time series. We also tested for heterogeneity in subsets of the data: considering only questions (rather than counting both questions and answers) and posts on weekdays. In both subsets our estimates did not deviate significantly from the main result: we estimate a 12\% relative decrease in questions and 14\% relative decrease in posts on weekdays.

\begin{table}[htbp]
    \begin{center}
        \label{main_results}
            \begin{tabular}{lccc} \hline
 & (1) & (2) & (3) \\
 & Number of posts & Number of questions & Weekday posts \\
&  &  &  \\
\hline
 &  &  &  \\
Stack Overflow $\times$ Post-GPT & -0.170** & -0.112+ & -0.149* \\
 & (0.0607) & (0.0619) & (0.0636) \\
 &  &  &  \\
Observations & 1,150 & 1,150 & 1,150 \\
R-squared & 0.995 & 0.994 & 0.993 \\
R-squared within & 0.290 & 0.315 & 0.232 \\
Outcome mean & 16363 & 7273 & 13191 \\
Outcome std. dev. & 29088 & 12661 & 23685 \\ \hline
\end{tabular}

    \end{center}
    \caption{Results of a difference-in-differences model, estimating the change in activity observed weekly on Stack Overflow following the release of ChatGPT, relative to activity on four other platforms less likely to have been impacted. All regressions comprise platform fixed effects, week fixed effects, and platform-specific linear time-trends. The standard-error of the estimate clustered on month is reported in parentheses. Significance codes: ***: $p< 0.001$, **: $p< 0.01$, *: $p<0.05$, +: $p<0.1$.} 
    \label{tab:diff_in_diff}
\end{table}

Figure~\ref{fig:interaction_plot} shows that the impact of ChatGPT is increasing over time and is by the end of our study greater in magnitude than the average post-ChatGPT effect estimated in Table~\ref{tab:diff_in_diff}. By the end of April 2023, the estimated effect stabilizes at around 25\%. Interestingly, ChatGPT use, in general, peaked around this time.\footnote{\url{https://www.similarweb.com/blog/insights/ai-news/chatgpt-bard/}}

\paragraph{Voting activity}
A decrease in overall activity on Stack Overflow does not necessarily signify a problem; it could indicate a beneficial shift toward fewer but higher quality posts, as less valued or simplistic questions may be outsourced to ChatGPT. We investigate this possibility using data on voting activity but observe no significant change in the typical appreciation of posts after ChatGPT's release.

\begin{figure}[H]
    \centering
    \includegraphics[width = 0.5\textwidth]{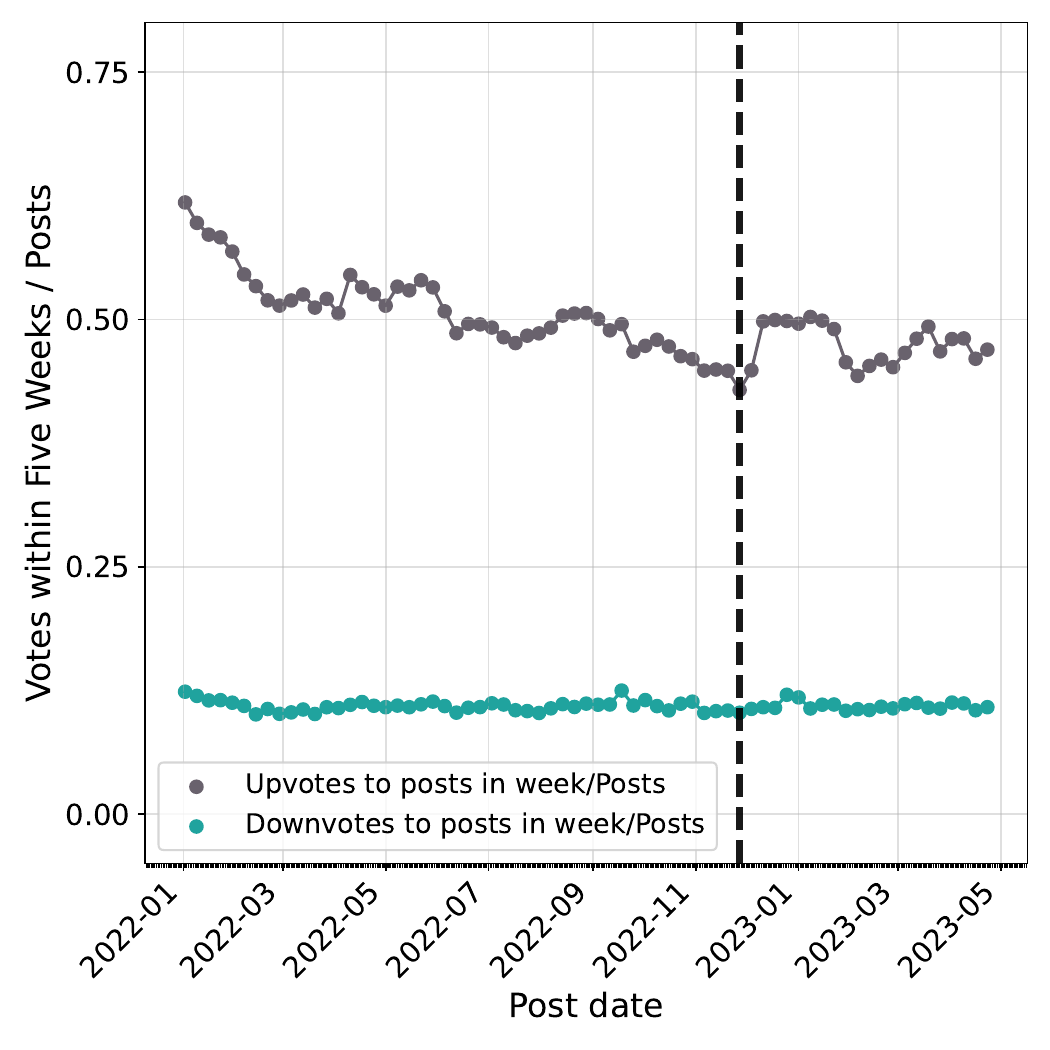}
    \caption{The time series of upvotes and downvotes accruing to posts within five weeks of their appearance. We observe no significant change since the release of ChatGPT. The horizontal axis indicates the week of the post.}
    \label{fig:votes}
\end{figure}

The time series of upvotes and downvotes, which we use as a proxy for the overall quality of posts, remain stable across the release of ChatGPT. Specifically, Figure~\ref{fig:votes} reports the average number of upvotes and downvotes that posts from a given week receive within five weeks of their creation. Upvotes are shown in grey and downvotes in blue; neither series changes significantly. Indeed the relative stability of voting behavior suggests that the quality of posts on Stack Overflow has not meaningfully changed after the introduction of ChatGPT.

\subsection{Heterogeneities across tags}
Studying posts about different programming languages on Stack Overflow, we find significant heterogeneities in the impact of ChatGPT on posting behavior across languages. 

\begin{figure}[H]
    \centering
    \includegraphics[width=\textwidth]{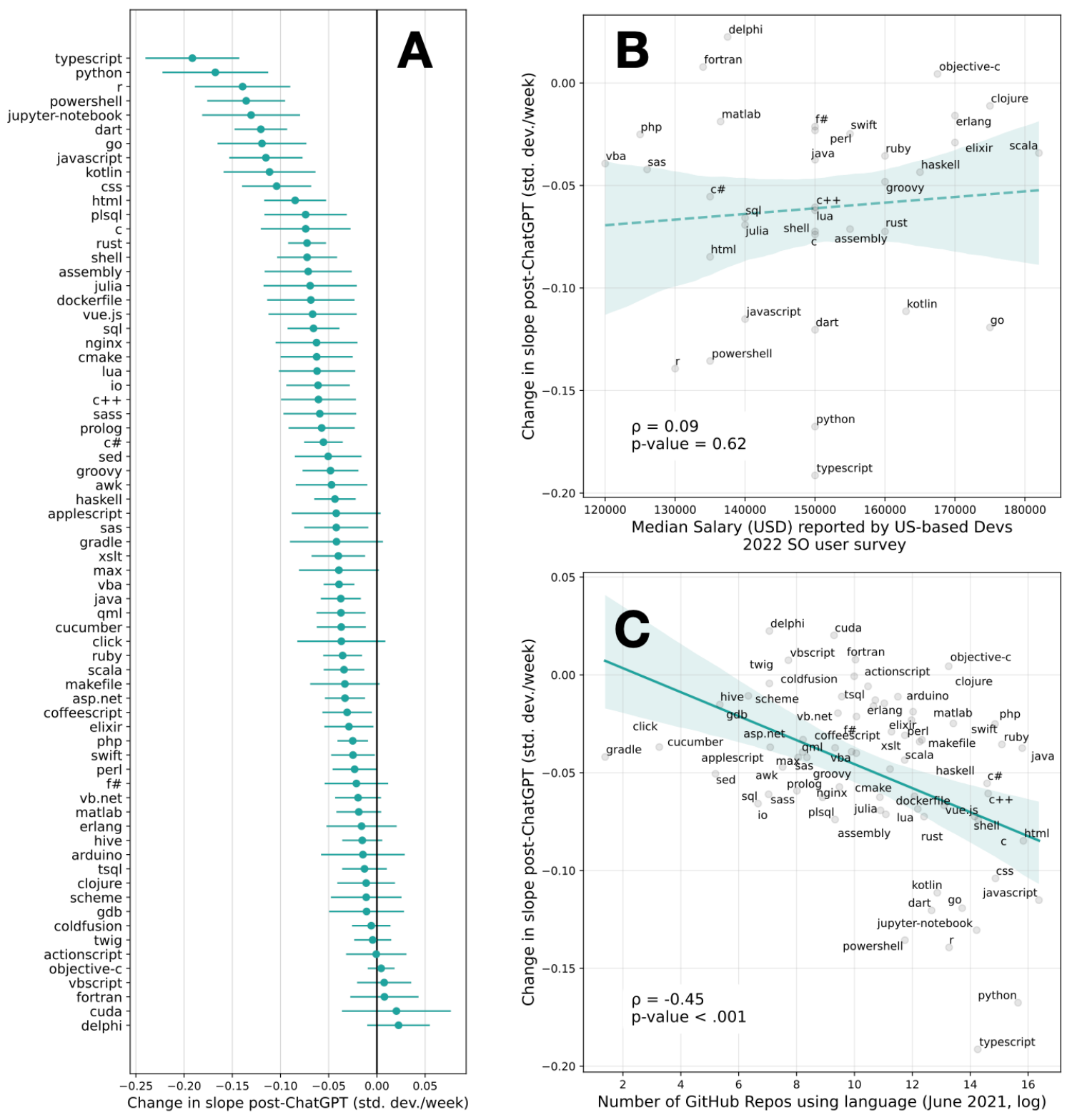}
    \caption{A) The event study estimates of the effect of ChatGPT's release on activity on a selection of tags on Stack Overflow. We report HAC-corrected 95\% confidence intervals. B) The relationship between estimated effects and salary data from the Stack Overflow developer survey. We find no significant relationship. C) The relationship between the number of GitHub repositories using a tag and the estimated effect of ChatGPT on that tag. In both B) and C) we plot a linear fit with bootstrapped 95\% confidence intervals. The dashed line in B) indicates that the correlation is not significant}
    \label{fig:tag_level}
\end{figure}

In Facet A of Figure~\ref{fig:tag_level}, we plot the estimated effects (slope changes in the linear time trend after the introduction of ChatGPT) for those 69 tags that we connected to a programming language on GitHub. We estimate a negative effect of ChatGPT for most tags, but the estimates range between a 0.25 standard deviation decrease in slope (i.e. change per week following the ChatGPT release) to a 0.03 standard deviation \textit{increase}. We observe that some of the widely used languages like Python and Javascript are the most impacted by ChatGPT. Interestingly, the model estimates that posts about CUDA have increased (though not significantly) after ChatGPT was released. CUDA is an application programming interface created by Nvidia, a graphics card manufacturer, that facilitates the use of graphics cards for computational tasks, in particular for machine learning and artificial intelligence. This exception again demonstrates the impact of ChatGPT on the world of computer programming: people are increasingly interested in software relating to artificial intelligence.

In Facet B, we compare the estimated impact of ChatGPT on different languages against salary data of developers using those languages. We source salary data from the 2022 Stack Overflow developer survey, focusing on US-based developers and calculating medians of reported salaries. We observe no clear relationship between the estimated labor market value of a specific language and changes in posting behavior in that language post-ChatGPT. 

To better understand the relationship between the size of the user base of a programming language and how it is impacted by ChatGPT, we compare our estimates with data from GitHub, the largest online platform for collaborative software development. Among other sources, ChatGPT was trained on data from GitHub. Because training data was collected up to September 2021, we use data on language use on GitHub up to June 2021. In Facet C of Figure~\ref{fig:tag_level}, we visualize the relationship between the number of GitHub repositories (coding projects) in a specific language and the estimated impact of ChatGPT on that language. We observe that languages with more GitHub repositories tend to be more significantly impacted by the release of ChatGPT in terms of associated activity on Stack Overflow (Pearson's $\rho = -0.45, p<.001$).

\section{Discussion}\label{sec:discussion}

The rate at which people have adopted ChatGPT is one of the fastest in the history of technology \citep{teubner2023welcome}. It is essential that we better understand what activities this new technology displaces and what second-order effects this substitution may have \citep{Schumpeter1942,aghion1992model}. This paper shows that after the introduction of ChatGPT there was a sharp decrease in human content creation on Stack Overflow. We compare the decrease in activity on Stack Overflow with other Stack Exchange platforms where current LLMs are less likely to be used. Using a difference-in-differences model, we find about $16\%$  relative decrease in posting activity on Stack Overflow, with a larger effect in later months. We observed no large change in social feedback on posts, measured using votes, following ChatGPT's release, suggesting that average post quality has not changed. Posting activity related to more popular programming languages decreased more on average than that for more niche languages. These results suggest that users partially substituted Stack Overflow with ChatGPT. Consequently, the wide adoption of LLMs can decrease the provision of digital public goods, in particular, the open data previously generated by interactions on the web.

Our results and data have some shortcomings that point to open questions about the use and impact of LLMs. First, while we can present strong evidence that ChatGPT decreased the posting activity in Stack Overflow, we can only partially assess quality of posting activity using data on upvotes and downvotes.
Users may be posting more challenging questions, ones that LLMs cannot (yet) address, to Stack Overflow. Future work should examine whether continued activity on Stack Overflow is more complex or sophisticated on average than posts from prior to ChatGPT release. Similarly, ChatGPT may have reduced the volume of duplicate questions about simple topics, though this is unlikely to impact our main results as duplicates are estimated to account for only $3\%$ of posts \citep{correa2013fit}, and we do not observe significant changes in voting outcomes. 

A second limitation of our work is that we cannot observe the extent to which Russian- and Chinese-language users of the corresponding Q\&A platforms are actually hindered from accessing ChatGPT; indeed recent work has shown a spike in VPN and Tor activity following the blocking of ChatGPT in Italy \citep{kreitmeir2023unintended}. Given the potential economic importance of ChatGPT and similar LLMs, it is anyway essential that we better understand how such bans and blocks impact the accessibility of these tools \citep{gaessler2023training}. Finally, we do not address the issue that ChatGPT may be used to generate Stack Overflow content. Stack Overflow policy effectively banned posts authored by ChatGPT within a week of its release. In any case, a significant amount of ChatGPT activity on Stack Overflow would mean that our measures underestimate the effect of ChatGPT.

Despite these shortcomings, our results have important implications for the future of digital public goods. Before the introduction of ChatGPT, more human-generated content was posted to Stack Overflow, forming a collective digital public good due to their non-rivalrous and non-exclusionary nature -- anyone with internet access can view, absorb, and extend this information, without diminishing the value of the knowledge. Now, this information is rather fed into privately owned LLMs like ChatGPT. This represents a significant and trending shift of knowledge from the public domain to the private ones. 

This observed substitution effect poses several issues for the future of artificial intelligence in general. The first is that if language models crowd out open data creation, they will be limiting their own future training data and effectiveness. The second is that owners of the current leading models have exclusive access to user inputs and feedback, which, with a relatively smaller pool of open data, gives them a significant advantage against new competitors in training future models. Third, the decline of public resources on the web would reverse progress made by the web toward democratizing access to knowledge and information. Finally, the consolidation of humans searching for information around one or a few language models could narrow our explorations and focus our attention on mainstream topics. We briefly elaborate on these points, then conclude with a wider appeal for more research on the political economy of open data and AI, and how we can incentivize continued contributions to digital public goods.

\paragraph{Training future models}
Our findings suggest that the widespread adoption of ChatGPT may make it difficult to train few iterations \citep{taleb2012antifragile}. Though researchers have already expressed concerns about running out of data for training AI models \citep{villalobos2022will}, our results show that the use of LLMs can slow down the creation of new data. Given the growing evidence that data generated by LLMs cannot effectively train new LLMs \citep{gudibande2023false,shumailov2023recursion,alemohammad2023selfconsuming}, modelers face the real problem of running out of useful data. If ChatGPT truly is a ``blurry JPEG'' of the web \citep{chiang2023chatgpt}, then, in the long run, it cannot effectively replace its most important input: data derived from human activity. The proliferation of LLMs has already impacted other forms of data creation: many Amazon Mechanical Turk workers now generate content (i.e. respond to surveys, evaluate texts) using ChatGPT \citep{veselovsky2023turk}.

\paragraph{Competition in the artificial intelligence sector}
A firm's early advantage in technological innovation often leads to significant market share \citep{arthur1989competing}. In our case, ChatGPT is simultaneously decreasing the amount of open training data that competitors could use to build competing models, while capturing a valuable private source of user data. There is also a growing concentration in tech driven by a shift from companies going public to acquisitions \citep{ederer2023great} -- indeed OpenAI is partially owned by Microsoft. These forces may lead to a compounding advantage for OpenAI. Though firms have long used the massive amounts of open data created by users of platforms like Wikipedia, Stack Overflow, GitHub, OpenStreetMap or Reddit to create products and capture value \citep{henzinger2004extracting,vincent2018examining,vincent2021deeper}, these products have not generally replaced those platforms.

\paragraph{Lost economic value}
Digital public goods generate value in many ways besides feeding LLMs and other algorithms. For instance, Wikipedia is an important source of information worldwide, but in developing countries, readers are more often motivated by intrinsic learning goals and tend to read articles in greater detail \citep{lemmerich2019world}. Unequal access to artificial intelligence may also compound inequalities in growth and innovation between countries \citep{gaessler2023training}. 

Digital public goods also provide direct value to the many websites that extract data from open data to complement their core services with extra information \citep{piccardi2021value}. For instance, there is substantial interdependence between sites like Wikipedia, Reddit, and Stack Overflow and the search engines that use them to enrich responses to user queries via infoboxes \citep{mcmahon2017substantial,vincent2021deeper}. 

Contributors to digital public goods like Stack Overflow or Open Source Software (OSS) often enjoy indirect benefits \citep{lerner2002some}. For instance, while OSS itself provides significant value in the global economy \citep{greenstein2014digital}, OSS contributions are valuable signals of a firm's capabilities to investors \citep{conti2021beefing}. Individual contributions to Stack Overflow are used to signal ability on the labor market \citep{xu2020makes}. Any general tendency of ChatGPT to crowd out contributions to digital public goods, may limit these valuable signals that reduce economic frictions. On the other hand, such signaling activity may serve as a powerful incentive to keep people contributing.

\paragraph{Narrowing of information seeking}
The substitution effect we report likely has important second-order effects on how people search for information and their exposure to new ideas. LLMs likely favor well-established perspectives and due to their efficiency decrease the need for users to forage for information. These features of LLMs may reinforce a trend observed earlier in the context of the web. Specifically, internet search engines are thought to have pushed science toward consensus and narrower topics by improving efficiency of information search and improving the visibility of mainstream information \citep{evans2008electronic}. LLMs may also disincentivize the use of new or niche tools because they most amplify our productivity with those tools for which it has much training data. For instance, ChatGPT may not be able to help users of a new programming language that is has not seen many examples of. Given that LLMs are poised to change how we do research \citep{grossmann2023ai} and present a strong competitor to search engines \citep{xu2023chatgpt}, we need to understand what LLM efficiency implies for our contact with diverse sources of information and incentives to try new things.

More generally, models like ChatGPT are going to generate political and economic winners and losers like many previous breakthrough technologies. While early evidence shows that these models enhance productivity especially among new and inexperienced workers \citep{noy2023experimental,brynjolfsson2023generative}, there are other ways in which they may contribute to inequality between people and firms \citep{rock2019engineering}, for instance via potential negative side effects of automation \citep{acemoglu2019automation,eloundou2023gpts}. Our results suggest that the economics of data creation and ownership will become more salient: as data becomes more valuable, there will be growing interest in how creators of data can capture some of that value \citep{li2023dimensions}. These multi-faceted aspects of the impact of LLMs suggest that the political economy of data and AI will be especially important in the next years \citep{lehdonvirta2022cloud,johnson2023power}.

In this context, our work highlights the specific issue that valuable digital public goods may be under-produced as a result of the proliferation of AI. A natural follow-up question is how we can incentivize the creation of such goods. While unemployment shocks are known to increase the provision of digital public goods \citep{kummer2020unemployment}, it would be an unsatisfying solution to suggest that people put out of work by automation will fill this gap. In the case of platforms like Stack Overflow, active users are often motivated by social feedback and gamification \citep{anderson2012discovering}, but the continual onboarding of new users is what keeps these platforms relevant in the long run \citep{danescu2013no}. For the sake of a sustainable open web and an AI ecosystem that draws on its data, we should think about how to keep people exchanging information and knowledge online.


\section{Appendix}\label{sec:methods}

\subsection*{Data}

\paragraph{Stack Exchange platform sites}
The raw dataset obtained from \url{https://archive.org/details/stackexchange} contains nearly all posting activity on the question and answer platforms hosted on the Stack Exchange network from its launch in 2008 to early June 2023. These include Stack Overflow, its Russian language version, and Math Overflow and Math Stack Exchange. Stack Overflow is the largest online Q\&A platform for topics relating to computer programming and software development. It  provides a community-curated discussion of issues programmers face \citep{anderson2012discovering}. Questions have multiple answers, and users debate the relative merits of solutions and alternatives in comments. A track record on Stack Overflow has value on the labor market as a signal of an individual's skills \citep{xu2020makes}. 

The data contains over 58 million posts, including both questions and answers. Posts are linked to their posting users, from which we infer poster previous activity and can identify posts made by new users. Questions are annotated with tags indicating the topic of the post including programming languages used. Users can give posts upvotes or downvotes, providing posting users with social feedback and reputation points. The Russian language version of Stack Overflow (over 900 thousand posts) and the mathematics-oriented platforms Math Stack Exchange (over 3.5 million posts) and Math Overflow (over 300 thousand posts) have identically structured data dumps hosted in the same location.

Registered users can upvotes and downvote posts made on Stack Exchange platforms. These votes provide a valuable signal of the value of posts \cite{anderson2012discovering,mamykina2011design}. They are the primary way users earn reputation points and status on Stack Exchange platforms. Votes also influence the ranking of posts in user feeds and search engine results, facilitating information filtering. Downvotes are used to moderate. The Stack Exchange data dump contains data on every vote cast, including the corresponding post, the date the vote was made, and whether it was an upvote or downvote.

\paragraph{Segmentfault}
Segmentfault is a Chinese language platform with a Q\&A platform for developers that has many similarities with the Stack Exchange sites. Users post questions on programming language topics and other users post answers. Questions are tagged by relevant languages and technologies, and there are similar gamification elements on the platform. We scraped data on all posts as of early June 2023, gathering over 300 thousand in total.

\paragraph{Selection of tags} 
Stack Overflow posts are annotated by tags which describe the concepts and technologies used in the post. For example, many tags indicate programming languages, webframeworks, database technologies, or programming concepts like functions or algorithms. Stack Overflow reconciles tags referring to the same things via a centralized synonym dictionary. We selected the 1,000 most used tags up to early June 2023, and focused on those 69 which could be directly linked to language statistics reported by GitHub, described next. 

\paragraph{GitHub data on programming language use}
We use data from the June 2021 GHTorrent data dump \citep{gousios2012ghtorrent} as a proxy measure for the amount of open data available for each programming language. The dataset reports which languages are used in each project or repository on GitHub. We simply count the number of repositories mentioning each language. We then link the languages with tags on Stack Overflow. As an alternative, we count the number of commits, elemental code contributions to repositories, to each repository, hence language. In the main paper we visualize the estimated effects of ChatGPT on specific tags that we can link to GitHub languages. We exclude some tags which refer to file formats or plain text, specifically: yaml, json, text, svg, markdown, and xml.

\subsection*{Data and Code availability}
Data and code to reproduce our analyses will be made available in a subsequent draft. The Stack Overflow data dump is available here: \url{https://archive.org/details/stackexchange}.

\subsection*{Acknowledgments} We thank Frank Neffke, Gerg\H{o} T\'oth, Christoffer Koch, S\'andor Juh\'asz, Martin Allen, Manran Zhu, Karl Wachs, L\'aszl\'o Czaller, and Helene Strandt for helpful comments and discussions.

\bibliographystyle{agsm}
\bibliography{references}

\end{document}